\documentclass{iopart}
\usepackage{iopams}
\usepackage{graphicx}
\usepackage{url,graphics,epsfig}
\usepackage{amssymb}
\usepackage[breaklinks=true,colorlinks=true,linkcolor=blue,urlcolor=blue,citecolor=blue]{hyperref}
\usepackage{tikz}
\usepackage{cite}

\usepackage{xcolor}
\begin{document}
\jl{2}
%
%+++++++++++++++++++++++++++++++++++++++++++++++++++++++++++++++++++++++++++++
%
%  Macro definitions
%
%+++++++++++++++++++++++++++++++++++++++++++++++++++++++++++++++++++++++++++++
\def\etal{{\it et al~}}
\def\newblock{\hskip .11em plus .33em minus .07em}
%+++++++++++++++++++++++++++++++++++++++++++++++++++++++++++++++++++++++++++++
%  End of Macro definitions
%
%+++++++++++++++++++++++++++++++++++++++++++++++++++++++++++++++++++++++++++++
%
%+++++++++++++++++++++++++++++++++++++++++++++++++++++++++++++++++++++++++++++
%
% Title of the paper
%
%+++++++++++++++++++++++++++++++++++++++++++++++++++++++++++++++++++++++++++++
%
\setlength{\arraycolsep}{2.5pt}             % use this for journal style

\title[Photoionization of Rb$^{2+}$ ions] {Photoionization of Rb$^{2+}$ ions in the valence energy region 37 eV -- 44 eV}

\author{Brendan M McLaughlin$^{1,2}\footnote[1]{Corresponding author, E-mail: bmclaughlin899@btinternet.com}$ and 
               James  F Babb$^{2}\footnote[2]{Corresponding author, E-mail: jbabb@cfa.harvard.edu}$}

\address{$^1$Centre for Theoretical Atomic, Molecular and Optical Physics,
                          School of Mathematics and Physics, The Old Physics Building, 
                          Queen's University Belfast, Belfast BT7 1NN, UK}

\address{$^2$Institute for Theoretical Atomic and Molecular Physics, 
                          Center for Astrophysics, Harvard \& Smithsonian,
                          Cambridge, MA 02138, USA}
%
%+++++++++++++++++++++++++++++++++++++++++++++++++++++++++++++++++++++++++++++
%
%              Abstract
%
%+++++++++++++++++++++++++++++++++++++++++++++++++++++++++++++++++++++++++++++
\begin{abstract}
Absolute photoionization cross sections for the Rb$^{2+}$ 
ion were recently measured at high resolution
over the energy range  37.31 eV --  44.08~eV,
with autoionizing Rydberg resonance series identified, using the photon-ion merged-beam setup 
at the Advanced Light Source (\textsc{als})  in Berkeley
{(Macaluso D A et. al.  J. Phys. B: At. Mol. Opt. Phys. {\bf 49} (2016) 235002; {\bf 50} (2017) 119501)}.
Detailed photon-energy scans taken at 13.5  $\pm$ 2.5 meV bandwidth 
illustrated multiple Rydberg resonance series 
associated with the ground and metastable states.  
Here we present theoretical cross section results obtained using the Dirac-Coulomb 
$R$-matrix approximation with a detailed analysis of the resonances. 
%utilising the time-delay of the $S$-matrix method. 
The calculations were performed for the $3d^{10}4s^2 4p^5 \; {^2}{\rm P^o}_{J}$, 
$J=\frac{3}{2}$ ground state and the corresponding
$3d^{10}4s^2 4p^5 \; {^2}{\rm P^o}_{J}$, $J=\frac{1}{2}$  metastable level. 
Results from the large-scale calculations are benchmarked against 
the  \textsc{als} high-resolution measurements and reproduce 
the dominant resonance features in the spectra, 
providing confidence in the theoretical work 
for astrophysical applications.
\end{abstract}


\vspace{0.25cm}
\begin{flushleft}
%Short title: Valence shell photoionization of the atomic rubidium  ion Rb$^{2+}$\\
Short title: Photoionization of the atomic rubidium ion Rb$^{2+}$\\
\vspace{0.25cm}
\submitto{\jpb: \today}
\end{flushleft}
\maketitle
%
%++++++++++++++++++++++++++++++++++++++++++++++++++++++++++++++++++++++++++++
%
%      Text of paper follows
%
%++++++++++++++++++++++++++++++++++++++++++++++++++++++++++++++++++++++++++++
\section{Introduction}
Photoionization of atomic ions is an important process in determining
the ionization balances and the abundances of elements
in photoionized astrophysical nebulae. 
It has recently become
possible to detect neutron-capture elements 
(Se, Cd, Ga, Ge, Rb, Kr, Br, Xe, Ba, Pb, {\color{black} etc.}) in 
ionized nebulae {\color{black}\cite{Pequignot1994,Sharpee2007,Sterling2008,Otsuka2013,Sterling2015,Sterling2016}}. 
These elements are produced by slow ($s$-process) 
or rapid ($r$-process)  neutron-capture nucleosynthesis. 
Measuring the abundances of these elements helps to reveal their dominant production
sites in the Universe, as well as details of stellar structure,
mixing and nucleosynthesis \cite{Sharpee2007,Lan2001,Sterling2016},
{\color{black} but abundance determinations are affected by
a lack of atomic data~\cite{Kwitter2014,Sterling2015b,Stasinska2017}.
In addition to what can be learned from
planetary nebulae, Walker \textit{et al.}~\cite{Walker2009} used observations
of the interstellar $^{85}\textrm{Rb}/^{87}\textrm{Rb}$ abundance ratio to gain insight
into the relative roles of various $n$-capture processes in massive stars.}
Such astrophysical observations are the motivation to determine the 
electron-impact excitation (EIE), electron-impact ionisation (EII), photoionization (PI), and
recombination properties of neutron-capture elements 
\cite{Card1993,Smith2015,Wall1997,Busso1999,Lan2001,Trav2004,Herwig2005,Sneden2008}.

{\color{black}With respect to Rb, specifically in support of astrophysical applications, 
recent work includes calculations of 538 transition energies, transition
probabilities, and oscillator strengths of $\textrm{Rb}^{3+}$, substantially
adding to the known data~\cite{Sansonetti2006,Persson1985},
and calculations of transition probabilities and collision strengths for 10 forbidden
transitions originating in the ground state of $\textrm{Rb}^{3+}$~\cite{Sterling2016}.
In addition, photoionization cross sections of
$\textrm{Rb}^{+}$ were measured~\cite{Macaluso2013,Kilbane2007,Neogi2003} 
and calculated~\cite{Babb2019,Costello2007}.
}

The motivation for the present study of the {\it trans}-Fe element Rb$^{2+}$
is to provide benchmark PI cross section data for applications in astrophysics,
{\color{black}to aid in the formulation of so-called ``ionization correction factors'' used
in the modeling of planetary nebular emission lines of 
ions of Rb~\cite{Shaw2015,Sterling2016,Stasinska2017}.}
High-resolution measurements of the photoionization cross section for Rb$^{2+}$  
with experimentally resolved Auger Rydberg resonance series
were recently published in the literature \cite{Macaluso2016,Macaluso2017}.
These high-resolution measurements, made over the photon energy range {\color{black}37.31 -- 44.08 eV} 
at the \textsc{als} synchrotron radiation facility in Berkeley, California, 
were taken with a band pass resolution of 13.5 $\pm$ 2.5 meV full width half maximum (FWHM).  
In tandem with the experiment, $R$-matrix calculations were performed in the intermediate coupling $jK$
Breit-Pauli approximation facilitating the identification of 
several highly excited Auger Rydberg resonance states. 

In the present investigation we use a fully relativistic approach within the Dirac-Coulomb 
$R$-matrix approximation  to calculate the cross sections for ground and metastable states.
The high resolution \textsc{als} photoionization cross sections are used to benchmark
the present PI cross sections and  Auger Rydberg resonances.
Excellent agreement is found, providing confidence in our theoretical data for various astrophysical applications.

This paper is structured as follows: Section~2 presents a brief outline of the theoretical work that was carried out. 
Section~3 presents comparisons of the calculated PI cross sections and Rydberg resonance series 
with the experimental data.  In section~4 we  present a discussion of these results.
Finally in section~5, conclusions are drawn from the present investigation.
%%##########################################################################################
%
%       Theory section of the paper
%
%##########################################################################################
\section{Theory}\label{sec:Theory}
High-resolution PI cross-section measurements require state-of-the-art theoretical methods
with relativistic effects \cite{Dyall1989,Grant2006,Grant2007}, in order  to obtain suitable
agreement with experiment.  
The present work employs an efficient parallel version~\cite{Ballance2006,Fivet2012}  of the
Dirac-Atomic {\textit R}-matrix-Codes (\textsc{darc})
\cite{Norrington1987,Wijesundera1991,darc,McLaughlin2012a,McLaughlin2012b} developed for treating
electron and photon interactions with atomic systems.  This suite continues to evolve
\cite{McLaughlin2015a,McLaughlin2015b,McLaughlin2016c,McLaughlin2017b,McLaughlin2018,McLaughlin2019}
in order to address ever  increasing expansions for the target
and collision models used in electron and photon impact with heavy 
atomic systems \cite{Mueller2017b,Smyth2017}.
The latest example of the application to photoionization is the recent comparison of experiment and 
theory for the Ca$^+$~\cite{Mueller2017c} and Rb$^+$ \cite{Babb2019} ions and the 
theoretical work on neutral Fe \cite{Smyth2018}.  Suitable agreement of the \textsc{darc}
photoionization cross-sections with high resolution measurements
performed at leading synchrotron light sources was obtained.

\subsection{Atomic structure}
The \textsc{\textsc{grasp}} code \cite{Dyall1989,Grant2006,Grant2007} was used to generate 
the residual Rb$^{3+}$ target wave functions employed in our collision work.  All orbitals were 
physical up to $n=3$, and in addition the 4$s$, 4$p$, $4d$, $5s$, $5p$ and $5d$ orbitals were included.
We began by performing an extended averaged level (\textsc{eal}) calculation for the $n=4$ orbitals 
and extended these calculations with the addition of the $n=5$ orbitals.  
All \textsc{eal} calculations were performed on the lowest 18 fine-structure levels of the residual 
Rb$^{3+}$ ion  in order to generate target wave functions for our photoionization studies. 
In our work we retained all the 687 - levels originating from one, and 
two--electron promotions from the $n=4$ levels into the orbital space of this ion. 
All  687 levels arising from the sixteen configurations 
were included in the \textsc{darc} close-coupling calculation, namely, 
the one-electron promotions,
$3d^{10}4s^24p^4$,  
$3d^{10}4s4p^5$,    
$3d^{10}4s4p^44d$, 
$3d^{10}4s4p^45s$,  
$3d^{10}4s4p^45p$,  
$3d^{10}4s4p^45d$,  
$3d^{10}4s^24p^34d$, 
$3d^{10}4s^24p^35s$,  
$3d^{10}4s^24p^35p$ and
$3d^{10}4s^24p^35d$.  
In addition we include the two-electron promotions,   
$3d^{10}4s^24p^24d^2$, 
$3d^{10}4s^24p^25s^2$  
$3d^{10}4s^24p^25p^2$, 
$3d^{10}4s^24p^25d^2$,
$3d^{10}4p^44d^2$, and 
$3d^{10}4p^45s^2$.  
This provides a suitable representation of the residual Rb$^{3+}$ ionic levels, 
as shown in Table~\ref{tab1}, where 
the energies of the lowest 9 levels of the residual Rb$^{3+}$ ion from the 687-level 
\textsc{\textsc{grasp}} calculations are compared to the 
values from the \textsc{nist} tabulations \cite{NIST2018} 
and from the \textsc{als} measurements \cite{Macaluso2016}.   
We note that the average percentage difference of our theoretical 
energy levels compared with the \textsc{nist} values is approximately 7\%.
%
% Table 1
%
\begin{table}
%\begin{flushleft}
\centering
\small
\caption{Comparison of Rb$^{3+}$ energy levels from the \textsc{grasp} code (present work)
         with \textsc{nist} tabulations \cite{NIST2018}
	     and \textsc{als} measurements \cite{Macaluso2016}.
	     Energies are with respect to the ground state and given in eV.  
         Percentage differences $\Delta (\%)$  are included for completeness.}
\label{tab1}
\begin{tabular}{llccccccc}
\br
Level         &  Configuration 	     	&  Term			&$J$		& \textsc{nist} 		 	&\textsc{grasp}		&\textsc{als}	& $\Delta_1^c$  	&$\Delta_2^{d}$ \\
		&				&				&		& Energy$^{\dagger}$ 	&Energy$^a$	&Energy$^b$			&	 		 &\\	
		&				&				&		& (eV)		 		&(eV)			& (eV)					&  ($\%$)		& (\%)\\	
\mr
 1  		& $3d^{10}4s^24p^4$ 	&  $\rm^3P$  		&2   		& 0.000000			&0			&0					&\,\;0.0		&\,\;0.0\\  
 2  		&				&				&1   		& 0.776188			&0.753086		&0.776 $\pm$ 0.010		&-3.1			&-3.0\\  
 3  		& 				&				&0   		& 0.861415			&0.897535		&--					&3.9 			&--\\    
\\
 4  		& $3d^{10}4s^24p^4$ 	&  $\rm ^1D$		&2     		& 2.152076			&2.418777		&2.152 $\pm$ 0.010		&12.4			&12.4\\ 
 \\
 5 		& $3d^{10}4s^24p^4$ 	&  $\rm ^1S$		&0     		& 4.761436			&6.016326		&4.771 $\pm$ 0.010		&26.4			&26.1\\ 
\\ 
 6 		&$3d^{10}4s4p^5$ 	&  $\rm ^3P^o$		&2     		& 16.735202			&17.660408		& --					&5.5	 		& --\\ 
 7 		&				&				&1    		& 17.310417			&18.245856		& --					&5.4 			& --\\ 
 8 		&				&				&0     		& 17.681250			&18.610682		& --					&5.3 			& --\\ 
\\
 9 		&$3d^{10}4s4p^5$ 	&  $\rm ^1P^o$		&1     		& 20.831188			&22.187881		& --					&6.5 			& --\\ 
\mr
\end{tabular}
\\
\begin{flushleft}
$^{\dagger}$\textsc{nist} Atomic Spectra Database tabulations  \cite{NIST2018}.\\
$^a$Present  \textsc{grasp} theoretical energies.\\
$^b$\textsc{als} experimental analysis \cite{Macaluso2016}.\\
$^c\Delta_1(\%)$ the percentage difference between the \textsc{grasp} and \textsc{nist}\cite{NIST2018} experimental values.\\
$^d\Delta_2(\%)$ the percentage difference between the \textsc{grasp} and \textsc{als} \cite{Macaluso2016} experimental values.\\
\end{flushleft}
\end{table}

\begin{figure*}
\includegraphics[width=\textwidth]{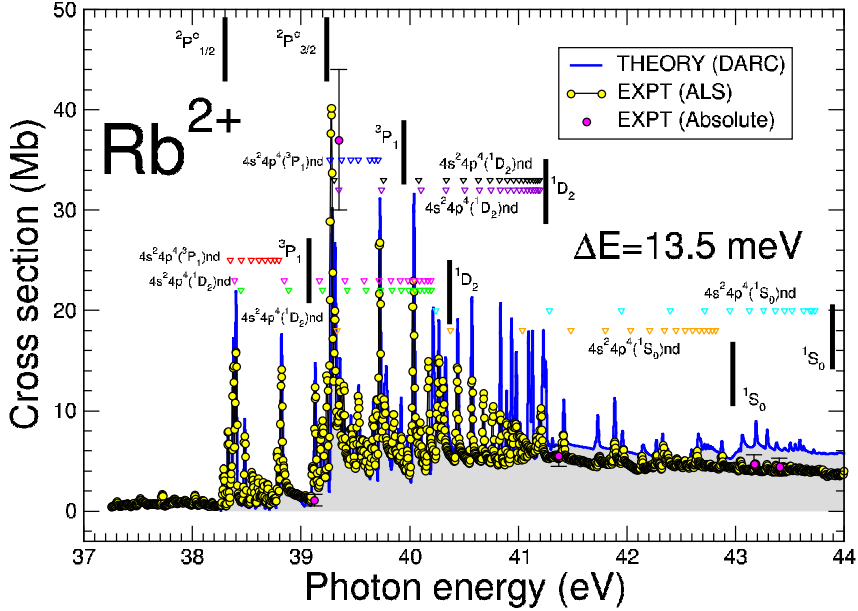}
\centering
\caption{(Colour online) 
                 Single photoionization cross section of Rb$^{2+}$ in the energy region 37 - 44 eV.
                 Experimental measurements (solid yellow circles) on an absolute scale 
	             with the absolute cross section measurements (solid magenta circles) made at selected energies
                 used to normalized the photo-ion yield spectra.
                 See the work of Macaluso and co-workers for further details \cite{Macaluso2016,Macaluso2017}.
                 Cross section results from the merged beam measurements at the \textsc{als} radiation facility,  
                 obtained  at a photon energy resolution of 13.5 $\pm$ 2.5 meV FWHM,
                  are compared with those from an 687-level Dirac $R$-matrix (\textsc{darc}) approximation.
                 The \textsc{darc} photoionization cross sections (solid blue line) were {\color{black} convolved} with a Gaussian
                 distribution having a 13.5 meV FWHM profile and statistically averaged 
                 for the  ground state  $3d^{10}4s^{2}4p^{5}\, {\rm ^2P^o}_{3/2}$ 
                 and $3d^{10}4s^{2}4p^{5}\, {\rm ^2P^o}_{1/2}$ metastable state (see text for details).
                 The inverted triangles are the energy positions of the various $4s^24p^4nd$ Rydberg resonance series 
                 found in the Rb$^{2+}$ photoionization spectrum, originating 
                 from the ${\rm ^2P^o}_{1/2}$ and ${\rm ^2P^o}_{3/2}$ initial states,
                 for the energy region investigated. {\color{black} The solid black vertical lines
                 indicate the photon energies required to ionize the Rb$^{2+}$  ${\rm ^2P^o}_{1/2}$ and ${\rm ^2P^o}_{3/2}$  states and
                 the limits for the each of the $4s^24p^4nd$ Rydberg series converging
                 to $^3\textrm{P}_1$, $^1\textrm{D}_2$, or $^1\textrm{S}_0$ states
                 of the residual Rb$^{3+}$ ion.}}
\label{fig1}
\end{figure*}
Photoionization cross section {\color{black} calculations} were  performed for the Rb$^{2+}$ ion
in the $3d^{10}4s^{2}4p^{5} \; {\rm ^2P^o}_{3/2}$ ground state  
and in the $3d^{10}4s^{2}4p^{5}\; {\rm ^2P^o}_{1/2}$  metastable level 
 using the \textsc{darc} codes with the above Rb$^{3+}$ residual ion  target wave functions. 

\subsection{Photoionization}
In our work we used sixteen continuum basis functions and a boundary 
radius of 14.65 Bohr to accommodate all the diffuse $n=5$ orbitals.
This resulted in Hamiltonian and dipole matrix sizes in excess of 45,000 
with over 3,000 coupled channels in our close coupling calculations.
For the ground and metastable initial states of the rubidium ions 
studied here, the outer region electron-ion collision problem was 
solved (in the resonance region below and  between all thresholds) using an extremely  
fine energy mesh of 5$\times$10$^{-8}$ Rydbergs ($\approx$ 0.680 $\mu$eV) 
  for the excited levels investigated. 
The $jj$-coupled Hamiltonian diagonal matrices energies were 
shifted  so that the theoretical term energies matched the recommended 
\textsc{nist} values~\cite{NIST2018}. We note that this energy adjustment ensures better 
positioning of resonances relative to all thresholds included in
the calculation \cite{McLaughlin2012a,McLaughlin2012b}.

In the present work the \textsc{darc} PI cross-section 
calculations  were convolved with a Gaussian distribution function
to simulate the experimental photon energy bandwidth.
Mathematically, a convolution is defined as the integral ${\cal I}$ over all space of 
the product of a function $F(x)$ and a function $G(u-x)$,
\begin{equation}
{\cal{I}} (u) = \int F(x) G(u-x) dx .
\end{equation}
The integration is taken over the variable $x$ (which may be a 1D or 3D variable), 
typically from minus infinity to infinity over all the dimensions. In the present work the
integration energy range for the convolution of the theoretical PI cross sections
 is that of the Rb$^{2+}$ experimental spectrum, 
 $F$ is the numerical theoretical cross section as a function of the photon energy 
and $G$ will be the Gaussian distribution \cite{Rohlf1994} corresponding to the resolution
of the experimental measurements.
Fig \ref{fig1} gives a comparison of the theoretical PI cross section from the 687-level  \textsc{darc} calculations 
with the \textsc{als} measurements.  In order to compare directly with experiment we have statistically averaged  over 
the ground and metastable initial states and {\color{black} convolved} our cross sections with a Gaussian distribution 
having a profile of 13.5 meV.  
The statistical averaged cross section $\sigma^{\rm Total}$ as a function of energy $E$ is 
\begin{equation}
\sigma^{\rm Total} (E) = \textstyle{}\frac{1}{3} \sigma({\rm ^2P^o_{1/2}})  + \frac{2}{3} \sigma({\rm ^2P^o_{3/2}}),
\end{equation}
with, respectively,  $\sigma({\rm ^2P^o_{3/2}})$  and $\sigma({\rm ^2P^o_{1/2}})$  the photoionization 
cross sections for the ground and metastable states.  Over the entire photon energy range 
investigated the theoretical results from the 687-level \textsc{darc} calculations are in good agreement 
with the high resolution \textsc{als} measurements.
As illustrated in Fig \ref{fig1}, multiple Rydberg resonance series are seen in the cross sections.
We now discuss their analysis.
\begin{figure*}
\includegraphics[width=\textwidth]{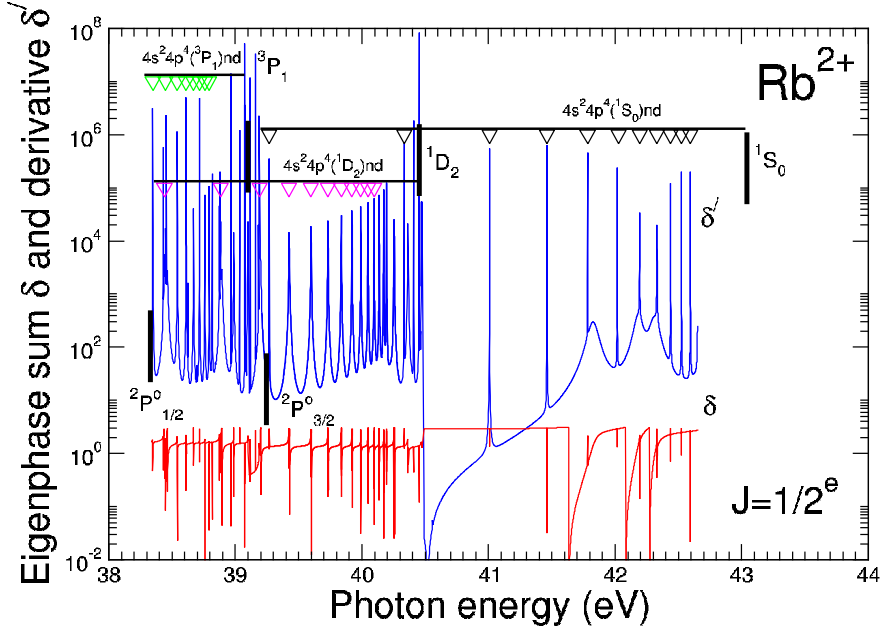}
\centering
\caption{(Colour online) 
                 Eigenphase sum ($\delta$, solid red line) and its derivative ($\delta^{\prime}$, solid blue line) for the J=1/2 
                 even scattering symmetry in the energy region 38 - 44 eV from the
                 \textsc{darc} photoionization out of the  $3d^{10}4s^{2}4p^{5}\, {\rm ^2P^o}_{1/2}$ metastable state
                 of the Rb$^{2+}$ ion.  The  prominent $4s^24p^4nd$  Rydberg Auger 
                 series parameters (found in the photoionization cross section) 
                 converging to the $\rm ^3P_1$, $\rm ^1D_2$ and $\rm ^1S_0$ thresholds of the Rb$^{3+}$ ion 
                 are tabulated in Table 2, 3 and 4. The inverted triangles are the 
                 energy positions of the various Rydberg resonance series.
                 Note, the presence of interloping resonances 
                 (assigned  to the $^1$S$_0$ threshold) perturb the 
                 regular Rydberg pattern lying below the $^1$D$_2$ threshold. 
                 In addition interloping resonances associated with higher lying 
                 thresholds perturb the Rydberg resonance series assigned to 
                 the $^1$S$_0$ threshold.}
\label{fig2}
\end{figure*}
\subsection{Resonances}
The multi-channel {\it R}-matrix eigenphase derivative
(QB) technique \cite{Burke2011}
as developed by Berrington and co-workers  \cite{qb1,qb2,qb3}
was used to locate and determine the resonance positions. 
The resonance autoionization width $\Gamma$ can be  determined from the inverse of the energy derivative
of the eigenphase sum $\delta$ \cite{Smith1960,Forrey1998} at the resonance energy $E_r$ via
\begin{equation}
	\Gamma = 2\left[ \frac{d\delta}{dE}\right]^{-1}_{E=E_r} = 2\left[\delta^{\prime}\right]^{-1}_{E=E_r} .
\end{equation}
The Auger widths of the resonances presented in Table \ref{tab2} - \ref{tab4} 
are determined via this method where the eigenphase sum is 
obtained from the large-scale multi-channel close-coupling 
electron scattering from the residual Rb$^{3+}$ ion within the context 
of the R-matrix method \cite{Burke2011}.

\begin{figure*}
\includegraphics[width=\textwidth]{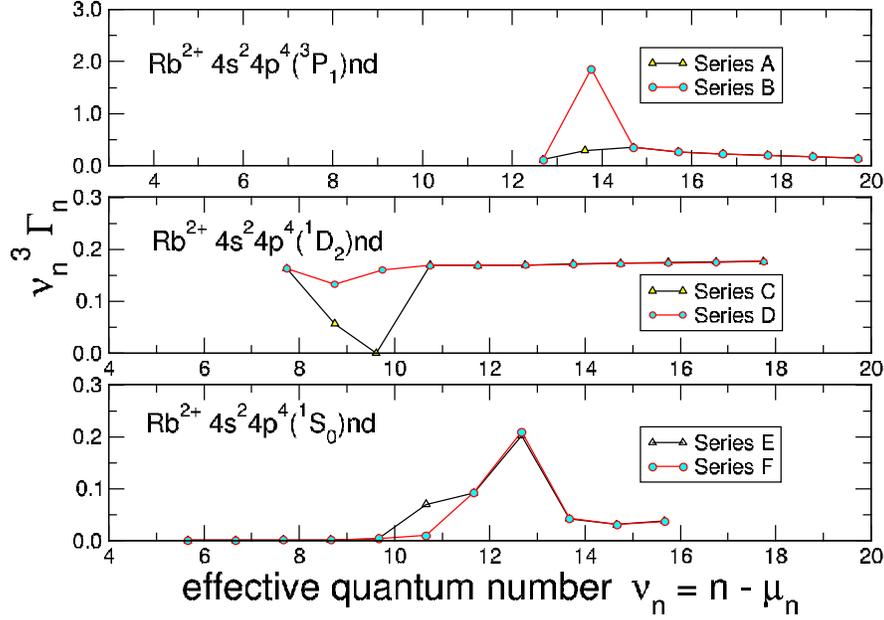}
\centering
\caption{(Colour online) 
                 {\color{black} The quantity $\nu_{n}^3\, \Gamma_{n}$ verses 
                 the effective quantum number $\nu_{n} = n - \mu_{n}$
                 for each of the six series, A - F, found in the spectrum.
                 The triangles are for the initial $J=\frac{1}{2}$ 
                 state and the corresponding circles are for the initial $J=\frac{3}{2}$
                 state. The top panel are the $4s^24p^4({\rm ^3P}_1) \ nd$ series going to the $\rm ^3P_1$ threshold,
                 the middle panel the $4s^24p^4({\rm ^1D}_2) \ nd$ series going to the $\rm ^1D_2$ threshold, and the 
                 bottom panel the $4s^24p^4({\rm ^1D}_2) \ nd$ series going to the $^1S_0$ threshold of the Rb$^{3+}$ ion.}}
\label{fig3}
\end{figure*}

Fig \ref{fig2} illustrates the eigenphase sum and its derivative, for the $J=\frac{1}{2}$ scattering symmetry, 
in the photon energy region 38 eV -- 44 eV, where prominent Rydberg series 
are found in the cross section for photoionization 
out of the initial metastable $^2P^o_{1/2}$ state of the Rb$^{2+}$ ion. 
As vividly seen from the behaviour of the eigenphase sum derivative, the presence of 
interloping  Rydberg resonances (associated with the $^1$S$_0$ threshold) 
perturb the regular resonance series and disrupt the regular Rydberg pattern 
lying below the $^1$D$_2$ and $^1$S$_0$ thresholds.  Such perturbating interlopers 
affect the widths of the resonances as is evidently seen from the 
results shown in Tables \ref{tab2} - \ref{tab4}, and in Fig \ref{fig3}.  
The QB method is suitable for determining the Auger widths of  isolated resonances.
{\color{black}We note that more elaborate approaches are
used to treat overlapping resonances~\cite{Shimamura2004,Shimamura2006,Shimamura2007,Shimamura2012}.}
{\color{black}In general, the eigenphase sum derivative, which is proportional
to the trace of the lifetime matrix, is expressible in the energy region
of resonance $n$ as
\begin{equation}
    d\delta/dE =\sum_n (\Gamma_n/2)/
    [(E_n-E)^2 + (\Gamma_n/2)^2] + d\delta_b/dE,
\end{equation}
where $d\delta_b/dE$ is the small background contribution.
The multi-channel $R$-matrix QB method used here, analyzes the eigenphase sum and its derivative 
to extract the resonance parameters. Results are presented in Table \ref{tab2} - \ref{tab4}. 
Overlapping resonances can in principle be analyzed by fitting
to this formula; however, when there are multiple
overlapping resonances a more comprehensive approach
is to examine the eigenvalues of the lifetime matrix,
see ~\cite{Shimamura2006,Shimamura2012} and 
sec.~3.2.4 of \cite{Burke2011}.}
For the present work the QB approach may be used to locate isolated resonance energies
and to highlight interloper effects in the spectra.

We attribute the difference in Auger widths to the different
number of coupled channels for the  $J=\frac{1}{2}$ and $J=\frac{3}{2}$ scattering symmetries. 
In addition the presence of overlapping and interloping resonances in the spectra 
disrupt the Rydberg patterns, which we analyze further in Sec.~\ref{Sec:discussion} below.

As shown in Tables \ref{tab2} - \ref{tab4}, theory 
and experiment are in harmony for the resonant 
energies and quantum defects. No measurements are available for the Auger 
widths due to the limited \textsc{als} experimental resolution.   
We note that all of the Auger widths $\Gamma$  for these Rydberg series
have a  natural line width very much less than 10 meV. 
In fact the bulk of them are less than 1 meV.  
Measurement of such narrow line widths are  beyond the present capabilities of the  \textsc{als} 
experimental resolution of 13.5 $\pm$ 2.5 meV.

An additional check on the theoretical data was carried out by 
calculating the integrated continuum oscillator strength $f$ from the experimental cross sections  
over the energy grid [$E_1$, $E_2$], where $E_1$ is the minimum experimental energy (37.31~eV) 
and $E_2$ is the maximum experimental energy (44.08~eV),  
using \cite{Shore1967,Fano1968,berko1979},
\begin{equation}
\label{eq:oscstr}
f    =  9.1075 \times 10^{-3} \int_{E_1}^{E_2} \sigma (h\nu) dh\nu   
%   & =  & 9.1075 \times 10^{-3} \; \; \overline{\sigma}_{\rm PI} ,  \nonumber \\
\end{equation}
Evaluating the continuum oscillator strength 
$f$ for the \textsc{als} cross section measurements yielded  a value of 0.294 $\pm$ 0.060,
where we have estimated an error of 20\% from the 
\textsc{als} absolute cross sections values \cite{Macaluso2016}. 
Evaluation of Eq.~(\ref{eq:oscstr}) for the theoretical Dirac {\it R}-matrix cross sections gave a
value of 0.420 for the statistical average of the 
$3d^{10}4s^{2}4p^{5} \; {\rm ^2P^o_{3/2}}$ ground state
and the $3d^{10}4s^{2}4p^{5} \; {\rm ^2P^o_{1/2}}$ metastable state.
This value is about 19\% outside the experimental error bar. 
The difference we attribute to the theoretical cross section being larger than 
experiment in the photon energy range 41.4 - 44 eV, 
as clearly illustrated in Fig \ref{fig1}.
We note that the Breit-Pauli {\color{black} cross section} calculations reported
in the work of Macaluso and co-workers \cite{Macaluso2016} lie consistently higher than experiment 
in this same photon energy range.
However, in their work Macaluso and co-workers \cite{Macaluso2016} 
did not report a value for the continuum oscillator strength $f$.
Experimentally, corrections for higher order radiation
effects \cite{Mueller2015} on the cross sections, or methods for piecing 
together all the various  high resolution scans to make a 
complete spectra (see \cite{Mueller2002,Schippers2003}),
might also be important.  However, in their work Macaluso and co-workers \cite{Macaluso2016} point out 
that higher order radiation effects are negligible above 40~eV in their Rb$^{2+}$ experiment 
contributing less than 1\% correction.

\section{Results and Discussion}\label{Sec:discussion}
As discussed above, 
{\color{black}the eigenphase sum derivative}
was used to extract the resonance parameters.
The resonance series identification can be made from Rydberg's
formula (see, for example, \cite{Seaton1983}) 
along with the \textsc{nist} tabulations \cite{NIST2018}.
The energy position $\epsilon_{\nu_n}$ of a resonance can be obtained via,
\begin{equation}
\epsilon_{\nu_n} = \epsilon_{\infty} - \frac{{\cal Z}^2}{\nu_n^2},
\end{equation}
with ${\cal Z}$ the charge of the core (in the present case 
${\cal Z}$ = 3)  and the effective quantum number $\nu_n=  n - \mu_n$, 
$\epsilon_{\infty}$ is the Rydberg series limit, $n$  is 
the principal quantum number and $\mu_n$ the quantum defect.
 
Table \ref{tab2} presents the Auger energies and quantum defects  for the Rydberg series found in 
 the $J=\frac{1}{2}$ and $J=\frac{3}{2}$ scattering symmetries, from the \textsc{darc} 687-level calculations, and
compares them with  the  measurements from the \textsc{als}.
The parameters are shown for the $4s^24p^4({\rm ^3P}_1) \ nd$ series converging to the  $\rm ^3P_1$ threshold 
of the Rb$^{3+}$ ion, where the initial state is the $4s^24p^5 ({\rm ^2P^o}_{3/2})$ ground state and the 
$4s^24p^5 ({\rm ^2P^o}_{1/2})$  metastable state. 
In Table \ref{tab3} results are shown  for the $4s^24p^4({\rm ^1D}_2) \ nd$ series converging 
to the  ${\rm ^1D}_2$ threshold and in  Table \ref{tab4} results 
are presented  for the $4s^24p^4({\rm ^1S}_0) \ nd$ series converging 
to the  ${\rm ^1S}_0$ threshold of the Rb$^{3+}$ ion.

First and foremost from Table \ref{tab2} - \ref{tab4} the widths of the various Rydberg resonance 
series are perturbed by the presence of interlopers. For regular Rydberg series then 
$\nu_n^3\Gamma_n$ should go smoothly to a constant 
as $\nu_n \rightarrow \infty$.  {\color{black}Note, 
from the results for the Auger widths presented in 
Table \ref{tab2} - \ref{tab4},  and in Fig \ref{fig3}},  
irregular behaviour is seen.  For the Auger widths 
presented in Tables \ref{tab2} -- \ref{tab4} 
the disruption of the smooth decay of the resonance width $\Gamma_n$ 
along the Rydberg series is due to the presence of interlopers.
This is evident for the resonance series associated with the ${\rm ^3P}_1$, 
${\rm ^1D}_2$ and ${\rm ^1S}_0$, Rb$^{3+}$ {\color{black} thresholds,} 
as illustrated in Fig~\ref{fig2}, {\color{black} and in Fig~\ref{fig3}}. 
For example, the Auger widths for the $n$ = 11, 12, 13, and 14 
members of the $4s^24p^4({\rm ^1S}_0) \ nd$ series, $J=\frac{1}{2}$ scattering 
symmetry, in Table \ref{tab4}, clearly illustrate this disruption effect. 
 	%
	%.   Table 2
	%
 \begin{table*}%[ht!]
 % \footnotesize
  \caption{Resonance energies $E_n$ and quantum defects $\mu_n$ of the $4s^24p^4({\rm ^3P}_1) \ nd$ Rydberg 
                series (from the 687-level \textsc{darc} calculations),  compared with the \textsc{als} 
                 measurements, for the $J=\frac{1}{2}$ and $J=\frac{3}{2}$ scattering symmetries, 
 		        converging to the Rb$^{3+}$($3d^{10}4s^24p^4\;{\rm ^3P}_{1}$) threshold originating from 
                 the Rb$^{2+}$ ground $4s^24p^5\;{\rm ^2P^o}_{3/2}$ and 
                 metastable state $4s^24p^5\;{\rm ^2P^o}_{1/2}$ . \label{tab2}}
\begin{center} 
 \begin{tabular}{ccclcccccc}
 
% \hline \hline \\  
\cline{1-1}                  	  	    \cline{2-4}     		 \cline{5-6}  				\cline{7-8}  	       	 \\
 %[-1.0ex]
 %    &&  \multicolumn{9}{c}{Autoionizing Rb$^{2+}$ Rydberg resonance series energies, J=1/2 symmetry}  \\ 
%[1.75ex]
 %                                          \cline{3-3}             \cline{7-7}         \\ 
[-1.0ex]
{\color{black} Series A}        	& &           $4s^24p^5({\rm ^2P^o}_{1/2})$  initial state 
	&&        	
	& &            $4s^24p^5({\rm ^2P^o}_{1/2})$  initial state            	
	& &                     				\\ 
[0.5ex]
                                           \cline{3-3}             \cline{7-7}     \\ 
        	& &           Experiment (\textsc{als})  
	&&       	
	& &            Theory (DARC)         	
	& &                     				\\ 
[0.5ex]                                           
%      	& & $4s^24p^4({\rm ^3P}_1) \ nd$
%	&& 	
%	& & $4s^24p^4({\rm ^3P}_1) \ nd$ 
%	& & 				\\ 
%[0.5ex]
% \hline \hline \\  
    \cline{1-1}     					\cline{2-4}                    \cline{5-6}  \cline{7-8}   	  \\ 
[-3ex]
   \\ 
[-1.75ex] 
  $n$   &$E_n$(eV)  & $\mu_n$ 	& 			&$E_n$(eV) & &$\mu_n$    	&$\Gamma_n$(meV)     &&\\ 
\cline{1-1}                  	  	    \cline{2-4}     		 	\cline{5-6}  				\cline{7-8}  	       	 \\
 13    	&38.352   	& 0.28    	 &			&38.348  & &0.31 		&0.8  			& &  \\
 14    	&38.459  	& 0.28     	&			&38.449  & &0.31  		&1.6   			&&     \\
 15    	&38.544  	&0.28  	&			&38.543  & &0.30    	&1.5  			&  &	 \\
 16    	&38.614 	& 0.28    	&			&38.612  & &0.30    	&0.9  			&   &\\
 17    	&38.671 	& 0.28     	&			&38.670  & &0.30    	&0.7   			&  &\\
 18    	&38.719 	&0.28      	&			 &38.718 &  &0.31    	&0.5  			& & \\
 19    	&38.760 	& 0.28    	&			&38.759 &  &0.30   		&0.4  			 &  &	\\
 20    	&38.794 	& 0.28 	&			&38.794 &  &0.28    	&0.3   			&    &\\			
 --	& -- 	        	& --	         &			& --  	    &  & --		&  	 		 &&	\\ 
\vdots &\vdots    	 & \vdots      &			&\vdots   & & \vdots   	&   			&& 	\\
  $\infty$ &39.109    &  --          &				&39.109  &  & --  	        	&   			&  &	\\ 
%\hline \hline                                                               
\cline{1-1}                  	  	    \cline{2-4}     		 \cline{5-6}  				\cline{7-8}  	       	 \\
 \\
 %[1.75ex]
 %                                          \cline{3-3}             \cline{7-7}         \\ 
[-1.0ex]
{\color{black}Series B}         	& &           $4s^24p^5({\rm ^2P^o}_{3/2})$  initial state 
	&&        	
	& &            $4s^24p^5({\rm ^2P^o}_{3/2})$  initial state            	
	& &                     				\\ 
[0.5ex]
                                           \cline{3-3}             \cline{7-7}      \\ 
     	& &           Experiment (\textsc{als})  
	&&       	
	& &            Theory (DARC)         	
	& &                     				\\ 
[0.5ex]                                           
%     & & $4s^24p^4({\rm ^3P}_1) \ nd$
%	&& 	
%	& & $4s^24p^4({\rm ^3P}_1) \ nd$ 
%	& & 				\\ 
%[0.5ex]
   \cline{1-1}     					\cline{2-4}                    \cline{5-6}  \cline{7-8}      \\ 
[-3ex]
   \\ 
[-1.75ex] 
  $n$   &$E_n$(eV)  	& $\mu_n$ 	&		&$E_n$(eV) 		&&$\mu_n$         		&$\Gamma_n$ (mev)	& &	\\ 
\cline{1-1}                  	  	    \cline{2-4}     		 \cline{5-6}  				\cline{7-8}  	       	 \\
 13    	&39.268   		& 0.27  	&		&39.263   		&&0.31			&0.8   		&   &	           \\
 14    	&39.374  		& 0.27      	&		 &39.379 		 &&0.32 			& 9.7  		&     &    	           \\
 15    	&39.459  		& 0.27 	&		 &39.457   		&&0.32    			& 1.5  		&      &    	 \\
 16    	&39.528 		& 0.27     	 &		 &39.526   		&&0.30    			&  0.9 		&       &  		  \\
 17    	&39.586 		& 0.27      	&		 &39.584  		 &&0.30    			&  0.7		 &       & 		 \\
 18    	&39.634 		& 0.27     	 &		&39.632   		&&0.32  			& 0.5 		& 	   &	  \\
 19    	&39.674 		& 0.27     	&		 &39.673  		& &0.30   			& 0.4  		&    &	   	\\
 20    	&39.709 		& 0.27 	 &		&39.707  		 &&0.30    			& 0.3 		 &    &	   	\\			
 --		& --	         & --	         	&		 &  --		 	&& --				&  --			 & 	& 	\\ 
\vdots 	& \vdots    &\vdots    	 &	 	&\vdots    		 && \vdots   			 & 			  &	&	 \\
  $\infty$ 	&40.023    &  --         	 &		&40.023   		 && --  	        		&   & &      		\\ 
\cline{1-1}                  	  	    \cline{2-4}     		 \cline{5-6}  				\cline{7-8}  	       	 \\                                                               
 \end{tabular} 
 \end{center}
 \end{table*}
 	%
	%.   Table 3
	%
 \begin{table*}%[ht!]
 % \footnotesize
  \caption{Resonance energies $E_n$ and quantum defects $\mu_n$ of the $4s^24p^4({\rm ^1D}_2) \ nd$  Rydberg 
           series (from the 687-level \textsc{darc} calculations),  compared with the \textsc{als} 
           measurements, for the $J=\frac{1}{2}$ and $J=\frac{3}{2}$ scattering symmetries, 
           converging to the Rb$^{3+}$($3d^{10}4s^24p^4\;{\rm ^1D}_{2}$) threshold originating from 
           the Rb$^{2+}$ ground $4s^24p^5\;{\rm ^2P^o}_{3/2}$ and 
           metastable state $4s^24p^5\;{\rm ^2P^o}_{1/2}$ . \label{tab3}}
\begin{center} 
 \begin{tabular}{ccclcccccc}
 
 %\hline \hline \\ 
\cline{1-1} \cline{2-4}  \cline{5-6}  \cline{7-8}  	       	 \\                     
\\
 % [-1.0ex]
 %   &&  \multicolumn{9}{c}{Autoionizing Rb$^{2+}$ Rydberg resonance series energies, J=1/2 symmetry}  \\ 
%[1.75ex]
 %                                          \cline{3-3}             \cline{7-7}         \\ 
[-1.0ex]
{\color{black} Series C}        	& &           $4s^24p^5({\rm ^2P^o}_{1/2})$  initial state 
	&&        	
	& &            $4s^24p^5({\rm ^2P^o}_{1/2})$  initial state            	
	& &                     				\\ 
[0.5ex]
                                           \cline{3-3}             \cline{7-7}        \\ 
        	& &           Experiment (\textsc{als})  
	&&       	
	& &            Theory (DARC)         	
	& &                     				\\ 
[0.5ex]                                           
%      & & $4s^24p^4({\rm ^1D}_2) \ nd$
%	&& 	
%	& & $4s^24p^4({\rm ^1D}_2) \ nd$
%	& & 				\\ 
%[0.5ex]
   \cline{1-1}     					\cline{2-4}                    \cline{5-6}  \cline{7-8}   	            \\ 
[-3ex]
   \\ 
[-1.75ex] 
  $n$   	&$E_n$(eV) 	& $\mu_n$ 		&		&$E_n$(eV) 		&&$\mu_n$         &$\Gamma_n$ (meV) &&	\\ 
\cline{1-1}                  	 \cline{2-4}      \cline{5-6}  \cline{7-8}             \\
  8    		&38.446   	& 0.25    		&			   	 &38.440   		&&0.26 	&4.8  	&&\\
 9    		&38.885  	& 0.25    		&			  	&38.884  		 &&0.25 	&1.2  	&& \\
 10    		&39.196  	& 0.25 		 &			 	&39.192  		 &&0.26  	& 10$^{-3}$ 	&& \\
 11    		&39.425 	& 0.25    		&			  	&39.424   		&&0.26    	&1.8   	&& \\
 12    		&39.598 	& 0.25    		&			 	&39.597   		&&0.26    	&1.4  	&&	 \\
 13    		&38.731 	&0.25    		&			   	&39.731   		&&0.26    	&1.1  	 &&\\
 14    		&39.837 	& 0.25    		&			  	&39.837   		&&0.26   	&0.9  	&& 	\\
 15    		&39.922 	& 0.25	 	&			 	&39.222   		&&0.26    	&0.7   	&& 	\\				
 16    		&39.991 	& 0.25	 	&			   	&39.991   		&&0.26    	&0.6  	&& 	\\			
 17    		&40.048 	& 0.25	 	&			   	&40.048   		&&0.26    	&0.5   	&& 	\\			
 18    		&40.096 	& 0.25	 	&			 	&40.096   		&&0.26    	&0.4  	&&	   	\\			
 --		& -- 	         	& --	         		&				& --  			&& --		&   		&&	 	\\ 
\vdots 	&\vdots     	&  \vdots         	&			  	 &\vdots    		&& \vdots   	&  		&&	  	\\
  $\infty$ 	&40.485    	&  --          		&			  	&40.485  		&& --  	        	&   		 &&  		\\ 
%\hline \hline                                                                \\
\cline{1-1}                  	  	    \cline{2-4}     		 \cline{5-6}  				\cline{7-8}  	       	 \\                                                               
\\
% [1.75ex]
 %                                          \cline{3-3}             \cline{7-7}         \\ 
[-1.0ex]
{\color{black}Series D}       	& &           $4s^24p^5({\rm ^2P^o}_{3/2})$  initial state 
	&&        	
	& &            $4s^24p^5({\rm ^2P^o}_{3/2})$  initial state            	
	& &                     			\\ 
[0.5ex]
                                           \cline{3-3}             \cline{7-7}        \\ 
        	& &           Experiment (\textsc{als})  
	&&       	
	& &            Theory (DARC)         	
	& &                     				\\ 
[0.5ex]                                           
%     	& & $4s^24p^4({\rm ^1D}_2) \ nd$
%	&& 	
%	& & $4s^24p^4({\rm ^1D}_2) \ nd$ 
%	& & 				\\ 
%[0.5ex]
%   \cline{1-1}     					\cline{2-3}                    \cline{5-6}  \cline{7-8}   	            \\ 
   \cline{1-1}     					\cline{2-4}                    \cline{5-6}  \cline{7-8}   	            \\ 
[-3ex]
   \\ 
[-1.75ex] 
  $n$   &$E_n$(eV)  	& $\mu_n$ 	&	&$E_n$(eV)	&&$\mu_n$       &$\Gamma_n$ (meV) &&\\ 
\cline{1-1}                  	  	    \cline{2-4}      \cline{5-6}  \cline{7-8}  	           \\
 8    		&39.347  	& 0.28     	&    	&39.355   	&&0.26	 &4.8  	 && \\
 9    		&39.790  	& 0.28    	& 	&39.797   	&&0.26  	& 2.7  	 && \\
 10    		&40.104  	&0.28  	&	&40.109 	&&0.26   	&2.4   	&& \\
 11    		&40.334 	& 0.28    	& 	&40.339  	&&0.26 	&1.9   	&&  \\
 12    		&40.508 	& 0.28    	 & 	&40.511   	&&0.26    	&1.4    	&&  \\
 13    		&40.643 	&0.28     	 &   	&40.645  	&&0.26   	&1.1		&& \\
 14    		&40.749 	& 0.28   	 &  	&40.751   	&&0.26   	&0.9   	&& 	\\
 15    		&40.834 	& 0.28 	 &   	&40.836   	&&0.26 	&0.7   	&& \\			
 16		&40.904	& 0.28      	 &	&40.905	&&0.26	&0.6   	&&	\\
 17		&40.961	& 0.28	 &	&40.962	&&0.26	&0.5   	&&	\\ 
 18		&41.009 	& 0.28	 &	&41.010	&&0.26	&0.4   	&&	\\ 
 --		& --	        	& --	        	 &	&  --		&&--		&   		&&	\\ 
 --		& --	        	& --	        	 &	&  --		&&--		&   		&&	\\ 
\vdots 	& \vdots    	&\vdots    	 & 	&\vdots    	& &\vdots    	&   		&&\\ 
  $\infty$ 	&41.399    	&  --          	& 	&41.399    	& &--  	        	&   		&&	\\ 
\cline{1-1}                  	  	    \cline{2-4}      \cline{5-6}  \cline{7-8}  	           \\                                                               
 \end{tabular} 
 \end{center}
 \end{table*}
 	%
	%.   Table 4
	%
		%
 \begin{table*}%[ht!]
 % \footnotesize
  \caption{Resonance energies $E_n$ and quantum defects $\mu_n$  of the $4s^24p^4({\rm ^1S}_0) \ nd$ Rydberg 
            series (from the 687-level \textsc{darc} calculations),  compared with the \textsc{als} 
           measurements, for the $J=\frac{1}{2}$ and $J=\frac{3}{2}$ scattering symmetries, 
 		   converging to the Rb$^{3+}$($3d^{10}4s^24p^4\;{\rm ^1S}_{0}$) threshold originating from 
            the Rb$^{2+}$ ground $4s^24p^5\;{\rm ^2P^o}_{3/2}$ and
           metastable state $4s^24p^5\;{\rm ^2P^o}_{1/2}$ . \label{tab4}}
\begin{center} 
 \begin{tabular}{ccclcccccc}
 
% \hline \hline \\  
\cline{1-1} \cline{2-4}      \cline{5-6}  \cline{7-8}  	% 
           \\
 %[-1.0ex]
 %   &&  \multicolumn{9}{c}{Autoionizing Rb$^{2+}$ Rydberg resonance series energies, J=1/2 symmetry}  \\ 
%[1.75ex]
%                                          \cline{3-3}             \cline{7-7}         \\ 
[-1.0ex]
{\color{black}Series E}        	& &           $4s^24p^5({\rm ^2P^o}_{1/2})$  initial state 
	&&        	
	& &            $4s^24p^5({\rm ^2P^o}_{1/2})$  initial state            	
	& &                     				\\ 
[0.5ex]
                                           \cline{3-3}             \cline{7-7}        \\ 
      	& &           Experiment (\textsc{als})  
	&&       	
	& &            Theory (DARC)         	
	& &                     				\\ 
[0.5ex]                                           
%      & & $4s^24p^4({\rm ^1S}_0) \ nd$
%	&& 	
%	& & $4s^24p^4({\rm ^1S}_0) \ nd$ 
%	& & 				\\ 
%[0.5ex]
   \cline{1-1}     						\cline{2-4}                    \cline{5-6}  \cline{7-8}   	            \\ 
[-3ex]
   \\ 
[-1.75ex] 
  $n$   	&$E_n$(eV)  	& $\mu_n$ 	&	&$E_n$(eV) 		&&$\mu_n$        &$\Gamma_n$ (meV) 	&&		\\ 
\cline{1-1}                  	  	    \cline{2-4}      \cline{5-6}  \cline{7-8}  	           \\
  6    		&39.332   	& 0.30     	&    	&39.267  	 	&&0.34	&0.1  			 &&  	           \\
  7    		&40.373  	& 0.30     	&	&40.333  	 	&&0.34 	&0.1   			       &&  	           \\
  8    		&41.037  	& 0.30    	& 	&41.008   		&&0.34 	&0.1  			        &&   	 \\
  9    		&41.484 	& 0.30     	&  	&41.462   		&&0.34    	&0.1   			        && 		  \\
 10    		&41.802 	& 0.30     	& 	&41.782   		&&0.34    	&0.1   			       &&	 \\
 11    		&42.034 	&0.30     	&    	&42.026 		&&0.34    	&0.8 			  &&	   	  \\
 12    		&42.209 	& 0.30     	&  	&42.194   		&&0.34  	&1.4   			 &&  	   	\\
 13    		&42.345 	& 0.30     	&   	&42.331   		&&0.34    	&0.2   			&&    	   	\\			
 14    		&42.451 	& 0.30     	&   	&42.438   		&&0.34    	&0.2   			 &&    	   	\\	
 15    		&42.537	& 0.30     	&   	&42.524  		&&0.34    	&0.1   			   &&	   	\\			
 16    		&42.607 	& 0.30    	&   	&42.595   		&&0.34    	&0.1 			   &&  	   	\\			
  --		& -- 	         & --	      	&	& --  			&& --		&  			  	&& 	\\ 
\vdots 	&\vdots     & \vdots         	 &	&\vdots    		&& \vdots   	&   				&&	  	\\
  $\infty$ 	&43.094    &  --          	&  	&43.094    		&& --  	        	&   			   &&   		\\ 
\cline{1-1}                  	  	    \cline{2-4}      \cline{5-6}  \cline{7-8}  	           \\
%\hline \hline                                                               
 \\
% [1.75ex]
%                                           \cline{3-3}             \cline{7-7}         \\ 
[-1.0ex]
{\color{black}Series F}    & &           $4s^24p^5({\rm ^2P^o}_{3/2})$  initial state 
	&&        	
	& &            $4s^24p^5({\rm ^2P^o}_{3/2})$  initial state            	
	& &                     				\\ 
[0.5ex]
                                           \cline{3-3}             \cline{7-7}        \\ 
         	& &           Experiment (\textsc{als})  
	&&       	
	& &            Theory (DARC)         	
	& &                     				\\ 
[0.5ex]                                           
%      	& & $4s^24p^4({\rm ^1S}_0) \ nd$
%	&& 	
%	& & $4s^24p^4({\rm ^1S}_0) \ nd$ 
%	& & 				\\ 
%[0.5ex]
   \cline{1-1}     					\cline{2-4}                    \cline{5-6}  \cline{7-8}   	            \\ 
[-3ex]
   \\ 
[-1.75ex] 
  $n$   	&$E_n$(eV)  & $\mu_n$ 	&	&$E_n$(eV) 		&&$\mu_n$    &$\Gamma_n$ (meV)     &&			\\ 
\cline{1-1}                  	  	    \cline{2-4}      \cline{5-6}  \cline{7-8}  	           \\
  6    		&40.242   	& 0.31     	 &    	&40.182  		& &0.34 	& 0.1	 &&	           \\
  7    		&41.286  	& 0.31     	&  	&41.248   		&&0.34 	& 0.1	 &&       	           \\
  8    		&41.950  	&0.31		& 	&41.922   		&&0.34	 &0.1	&&       	 \\
  9    		&42.398 	& 0.31   	&  	&42.376  		 &&0.34    	&0.1      	&&      		  \\
 10    		&42.715 	& 0.31     	& 	&42.697   		&&0.34    	&0.1       	&&     		 \\
 11    		&42.947 	& 0.31    	 &   	&42.931   		&&0.34    	&0.1  	 &&  	  \\
 12    		&43.127 	& 0.31   	&  	&43.108   		&&0.34   	&1.0      	&&	   	\\
 13    		&43.258 	& 0.31 	 &   	&43.245   		&&0.34    	&1.4   	 &&  	   	\\			
 14    		&43.365 	& 0.31 	 &  	&43.353  		 &&0.34    	& 0.2     	&& 	   	\\				
 15    		&43.451 	& 0.31 	 &   	&43.439   		&&0.34    	& 0.2   	 &&  	   	\\			
 16    		&43.521 	& 0.31 	 &  	&43.509 		 &&0.34    	&0.1     	&&  	   	\\			
 --		& --	         	& --	        	 &	&  --	      		 && --		&    		&& 	\\
\vdots 	& \vdots    	&\vdots     	&   	&\vdots     		&& \vdots    	&   		&&	 \\
  $\infty$ 	&44.088    	&  --          	&  	&44.088   		 && --  	 &      		&&   		\\ 
\cline{1-1}                  	  	    \cline{2-4}      \cline{5-6}  \cline{7-8}  	           \\
%\hline \hline                                                               
 \end{tabular} 
 \end{center}
 \end{table*}
\section{Summary and Conclusions}\label{sec:Conclusions}
We have carried out large-scale PI cross section calculations using the 
parallel version of the \textsc{darc} codes.  
Our statistically averaged cross sections
for the ground and metastable states show excellent agreement with the 
recent measurements from the \textsc{als} 
\cite{Macaluso2016} radiation facility from thresholds to about 41.4 eV.
The present theoretical cross sections in the photon region 41.4 - 44 eV 
are on average about 20\% higher than experiment,
but they are consistent with previous Breit-Pauli 
calculations performed on this complex \cite{Macaluso2016}. 
An analysis of the Auger Rydberg resonances series using the 
{\color{black}eigenphase sum derivative}
approach show excellent agreement with 
previous results {\color{black}within the experimental resolution}~\cite{Macaluso2016,Macaluso2017}.
Comparison between theory and  experiment 
for resonance energies and quantum defects 
in Tables \ref{tab2} - \ref{tab4}, provide further confidence 
in our theoretical data for applications in astrophysics. 
Numerical values of the DARC PI cross sections are available 
on request from the authors.

\clearpage
\ack
 B M McLaughlin acknowledges support
by the National Science Foundation (USA) through a grant to \textsc{itamp} at the 
Harvard-Smithsonian Center for Astrophysics under the visitors program,
 the University of Georgia at Athens for the award 
of an adjunct professorship and Queen's University Belfast
for a visiting research fellowship (\textsc{vrf}). 
Professor David Macaluso is thanked for the 
provision of the {\color{black}published} ALS data 
in numerical format and Captain Thomas J. Lavery USN Ret. 
for his constructive comments that enhanced the quality of this manuscript. 
 We dedicate this work to the staff of the cardiac unit at the \textsc{inova} hospital in Fairfax, 
 Virginia, USA, who graciously provided facilities for the completion of this manuscript.  
The authors acknowledge this research used grants of computing time at the National
Energy Research Scientific Computing Centre (\textsc{nersc}), which is supported
by the Office of Science of the U.S. Department of Energy
(\textsc{doe}) under Contract No. DE-AC02-05CH11231.
The authors gratefully acknowledge the Gauss Centre for 
Supercomputing e.V. (www.gauss-centre.eu) 
for funding this project by providing computing time on the GCS Supercomputer 
\textsc{hazel hen} at H\"{o}chstleistungsrechenzentrum Stuttgart (www.hlrs.de).
ITAMP is supported in part by NSF Grant No.\ PHY-1607396.
%+++++++++++++++++++++++++++++++++++++++++++++++++++++++++++++++++++++++++++++
%
%   Reference section now follows
%
%   Delete or change fake bibitem. delete next three
%   lines and directly read in your .bbl file if you use bibtex.
%
%+++++++++++++++++++++++++++++++++++++++++++++++++++++++++++++++++++++++++++++
%
\section*{References}

\bibliographystyle{iopart-num}

\bibliography{rb2plus}
\end{document}